# GRAVITAS : General Relativistic Astrophysics VIa Timing And Spectroscopy

**Kirpal Nandra**
*Max Planck Institute for Extraterrestrial Physics, 85741 Garching, Germany*

**Didier Barret**
*Institut de Recherche en Astrophysique et Planétologie, 9, Avenue du Colonel Roche, BP 44346, 31028, Toulouse Cedex 4, France*

**Andy Fabian**
*Institute of Astronomy, Madingley Road, Cambridge CB3 0HA, United Kingdom*

**Lothar Strueder**
*Max Planck Institute for Extraterrestrial Physics, 85741 Garching, Germany*

**Richard Willingale**
*Department of Physics and Astronomy, University of Leicester, Leicester LE1 7RH, United Kingdom*

**Mike Watson,**
*Department of Physics and Astronomy, University of Leicester, Leicester LE1 7RH, United Kingdom*

**Peter Jonker**
*SRON Netherlands Institute for Space Research, Sorbonnelaan 2, 3584 CA Utrecht, The Netherlands*

**Hideyo Kunieda**
*Division of Particle and Astrophysical Science, Graduate School of Science, Nagoya University, Furo-cho, Nagoya 464-8602, Japan*

**Giovanni Miniutti**
*Centro de Astrobiologia (CSIC-INTA), Centro de Astrobiologia (CAB), ESA - European Space Astronomy Center (ESAC), P.O. Box 78, 28691 Villanueva de la Cañada, Madrid, Spain*

**Christian Motch**
*Observatoire Astronomique de Strasbourg, 11 Rue de l'Université, F-67000 Strasbourg, France*

**Peter Predehl**
*Max Planck Institute for Extraterrestrial Physics, 85741 Garching, Germany*

# Abstract

GRAVITAS is an X-ray observatory, designed and optimised to address the ESA Cosmic Vision theme of "Matter under extreme conditions". It was submitted as a response to the call for M3 mission proposals. The concept centres around an X-ray telescope of unprecedented effective area, which will focus radiation emitted from close to the event horizon of black holes or the surface of neutron stars. To reveal the nature and behaviour of matter in the most extreme astrophysical environments, GRAVITAS targets a key feature in the X-ray spectra of compact objects: the iron Kα line at ~6.5 keV. The energy, profile, and variability of this emission line, and the properties of the surrounding continuum emission, shaped by General Relativity (GR) effects, provide a unique probe of gravity in its strong field limit. Among its prime targets are hundreds of supermassive black holes in bright Active Galactic Nuclei (AGN), which form the perfect laboratory to help understand the physical processes behind black hole growth. Accretion plays a fundamental role in the shaping of galaxies throughout cosmic time, via the process of feedback. Modest (~sub-arcmin) spatial resolution would deliver the necessary sensitivity to extend high quality X-ray spectroscopy of AGN to cosmologically-relevant distances. Closer to home, ultra-high count rate capabilities and sub-millisecond time resolution enable the study of GR effects and the equation of state of dense matter in the brightest X-ray binaries in our own Galaxy, using multiple probes, such as the broad iron line, the shape of the disk continuum emission, quasi-periodic oscillations, reverberation mapping, and X-ray burst oscillations. The enormous advance in spectral and timing capability compared to current or planned X-ray observatories would enable a vast array of additional scientific investigations, spanning the entire range of contemporary astrophysics from stars to distant galaxy clusters. Despite its breakthrough capabilities, all enabling technologies for GRAVITAS are already in a high state of readiness. It is based on ultra light-weight X-ray optics and a focal plane detector using silicon technology. The baseline launcher would be a Soyuz-Fregat to place GRAVITAS into a zero inclination, low-earth orbit, providing low and relatively stable background.

# 1 Science rationale

Almost a hundred years after the birth of GR, black holes remain among its most exciting and enigmatic predictions. Not only are black holes the manifestation of the strongest gravitational fields in the Universe, we now know that their growth can also have a profound effect on the formation and evolution of galaxies. Despite their extreme nature, only two parameters describe black holes in an astrophysical context: mass and spin. While black hole masses can be measured via the dynamics of relatively distant material, the effects of spin only manifest in the very innermost regions, where strong gravity dominates. Neutron stars are almost as extreme as black holes in terms of gravity, and contain the densest form of observable matter, in a regime where theory predicts the existence of exotic material such as hyperons and strange matter. The extreme limit of GR and the behaviour of matter at the highest densities cannot be investigated in a terrestrial environment. Black holes and neutron stars therefore provide unique laboratories to study processes critical not only to astrophysics, but also to cosmology and fundamental physics. Because the near environments of these compact objects are best studied via their X-ray emission, space-borne instrumentation is required. Here, space science can make a unique contribution to the advancement of physical knowledge.

GRAVITAS is a powerful facility designed specifically to address the theme of "Matter Under Extreme Conditions" of Cosmic Vision (Q3.3). It achieves this by observing the X-ray emission of accreting matter orbiting immediately around the event horizon of black holes and neutron stars. The continuum emission and characteristic spectral features, particularly of iron, reveal the effects of strong gravity through timing and spectroscopy. With a collecting area more than 20 times that of XMM-Newton EPIC-pn at the iron emission energy, GRAVITAS can detect variations of these features on timescales as short as the light crossing time at the event horizon, opening up the innermost region of black holes and neutron stars to intensive study.

GRAVITAS would reveal how general relativistic astrophysics works in the environments of black holes and neutron stars. Testing GR to high precision is not the main aim, rather it is to confirm whether or not matter behaves as expected in the strong gravity regime. It is only by studying how accretion works in detail that we can be secure in our understanding of astrophysical basis behind how quasars operate, and so affect the evolution of galaxies. Current observations with XMM-Newton have demonstrated the potential of this work, with time-averaged X-ray spectra and variability observations of a few bright objects. The next step to open up the time domain for typical objects, pushing below the timescale at the event horizon. This dictates an order of magnitude or more increase in collecting area around the iron line at ~6.4 keV compared to current facilities (*Figure 1*).

Our current picture of a luminous accreting black hole is of matter flowing through a disc liberating gravitational energy through radiation and magnetic energy processes, which energises a central corona. A hard power-law continuum irradiates the flow, giving rise to a back-scattered reflection spectrum, rich in atomic emission lines, particularly of iron. These form a powerful and unique diagnostic tool with which to study the geometry, velocity field and physical dimensions of the innermost region around the black hole or the neutron star, as well as the spin of the black hole through its proximity to the event horizon. It enables us to map regions of nanoarcsecond angular scale or smaller.

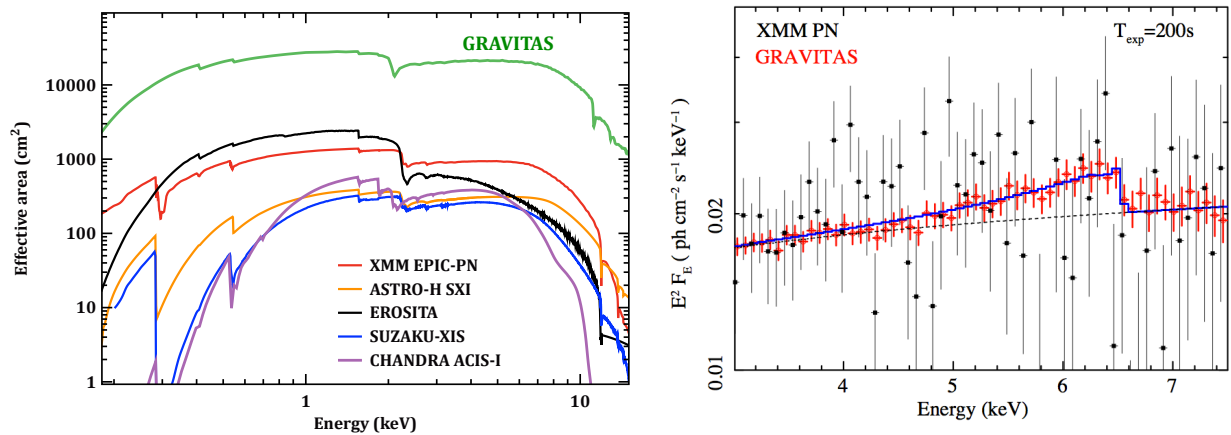

*Figure 1: Left) Comparison of the effective area of GRAVITAS to current and approved X-ray facilities. New lightweight optics technologies enable a huge leap forward in collecting area. Right) To illustrate the power of the mission for strong gravity work we compare the iron Kα profile of the archetypal AGN MCG-6-30-15 measured by EPIC-pn and GRAVITAS. The exposure is 200s, around the dynamical timescale at the black hole event horizon. GRAVITAS would open up this strong gravity regime around black holes via spectral timing.*

The equation of state of the densest matter in the Universe, found in the core of neutron stars, can be constrained using multiple probes with GRAVITAS. The key observable is the mass-radius (*M-R*) relation, for which different fundamental physics theories predict different forms. Reflection signatures from accreting neutron stars can yield measurements of their physical radii, as they can for black holes. Other variable and/or transient features of accreting neutron stars also become accessible, such as from bursts, quasi-periodic oscillations etc., which provide independent tests of the structure of these densest objects. Each of these methods yields constraints e.g. on $R$, $M/R$ or $M/R^2$, with an accuracy that for the first time would enable decisive discrimination between models.

Such a powerful telescope enables much additional science on a wide range of objects, from stars to distant clusters of galaxies (see Figure 2). It has the potential to make breakthroughs in a broad range of astrophysical investigations, from stars to distant clusters of galaxies. An an example, the properties of galaxy clusters and groups have important implications for cosmology via the growth of structure, the history of heavy element production, and the astrophysical evolution of hot gas. GRAVITAS would enable a major breakthrough in these fields by providing a large sample of high quality X-ray spectroscopic data for clusters and groups over a wide range of redshifts. Similarly, on the astrophysics of the X-ray background, GRAVITAS would measure the distribution of absorbing columns in the entire AGN population and provide measurements of the obscuration (and hence also intrinsic luminosity), of Compton thick AGN out to at least $z\sim1$; key for understanding the physical origin of the obscuration. Finally, GRAVITAS could be operating at an epoch where the focus of astronomy may well be shifting from large-scale imaging surveys (mostly addressing the issue of dark energy) to that of time domain. Major facilities such as LOFAR, LSST, and SKA will open up a new window on variable and transient phenomena, and the compact objects that produce them. GRAVITAS could provide the perfect complement in space to these facilities, with the power to discern the detailed astrophysics associated with the most extreme phenomena in the Universe with unprecedented precision.

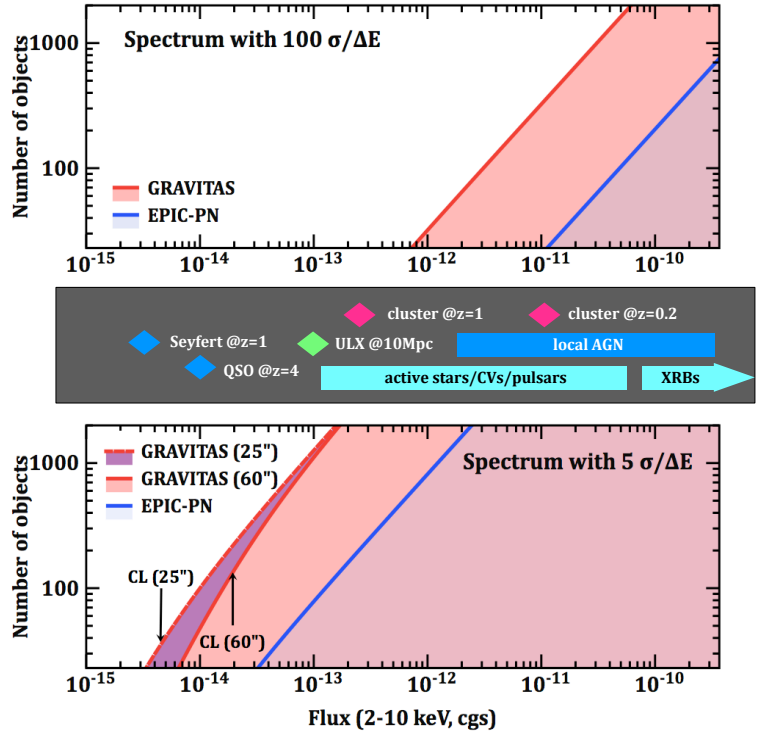

Figure 2: Performance of GRAVITAS for X-ray spectral determination compared with XMM-Newton EPIC-pn. Plots show the performance in terms of total number of objects per year that would be observable as a function of their hard band X-ray flux if the whole programme was devoted to their study, taking into account the expected/actual observing efficiencies. Top panel: performance for **very high quality spectra** with a statistical accuracy of $100\sigma$ per energy resolution element at 6.4 keV, corresponding to ~1,250,000 photons in the whole spectrum. Bottom panel: performance for **standard quality spectra** with a statistical accuracy of $5\sigma$ per energy resolution element at 6.4 keV, corresponding to ~3000 photons in the whole spectrum. In the bottom panel the confusion limits (CL) are shown for the 60" and 25" PSFs. The centre panel indicates the typical fluxes for some selected potential Galactic and extragalactic targets.

To achieve the GRAVITAS scientific objectives, the following science requirements have been derived (Table 1).

| Parameter | Value | Main drivers |
|---|---|---|
| **Effective area** | 1.5 m² @ 6.4 keV (goal 2m²) | Measure iron line parameters on sub-orbital timescales in AGNs |
| | 0.5 m² @ 10 keV | Constrain reflection continuum |
| | 1 m² @ 0.5 keV | NS atmospheres; BHB thermal disk |
| **Spectral resolution** | 125 eV @ 6.4 keV | Resolve broad iron Kα profile and deconvolve from narrower features |
| **Time resolution** | 100 µs (goal 16 µs) | Measure high frequency variations in NSs and BHBs |
| **Pile-up** | < 2% @ 150000 cps (0.3-15 keV)* | Measure spectral shape accurately |
| **Absolute timing accuracy** | 10 µs | Absolute phase comparison of ms pulsar folded light curves |
| **Throughput** | > 70% @ 150000 cps | Cope with the brightest X-ray sources |
| | > 50% @ 1 Mcps | |
| **Maximum count rate** | 1.5 Mcps | Observe type I X-ray bursts and X-ray novae in outbursts |
| **Angular resolution** | 1' (goal 25") | Avoid confusion for high z AGNs and clusters |
| **Attitude reconstruction (*a posteriori*)** | 12", 3σ (goal 5") | Accurate celestial location of serendipitous discoveries. |
| **Fast repointing capability** | 6 hours | Observe transient phenomena from external triggers |

Table 1: GRAVITAS science requirements.

## 2 The GRAVITAS space segment

The GRAVITAS space segment consists of a single spacecraft that includes a telescope assembly, a service module and an instrument module connected by an extendable optical bench (EOB). The telescope assembly is made of 6 identical telescopes with a focal length of ~12 meters. The focal plane instrumentation of GRAVITAS includes 6 identical HIgh Framerate Imagers (HIFIs). The HIFIs are protected from visible and X-ray stray-light using a conical baffle in front of each detector and from background particles (electrons, protons and cosmic-rays) using a magnetic diverter.

### 2.1 Optical configuration

*The telescopes must fit into the fairing of the Soyuz-Fregat launcher with its clear diameter of 3.8 m. In order to meet the required effective area of 1.5 m² at 6.4 keV, the telescopes must be arranged side-by-side in a way that minimizes the lateral separation between adjacent telescopes. Several possible implementations have been studied for the proposal and led to similar designs for the overall mission. Ray tracing of a 6 and 7 mirror configurations led to the effective area curves shown in*

Figure 3. The more conservative *six telescope* option was selected as the baseline for GRAVITAS.

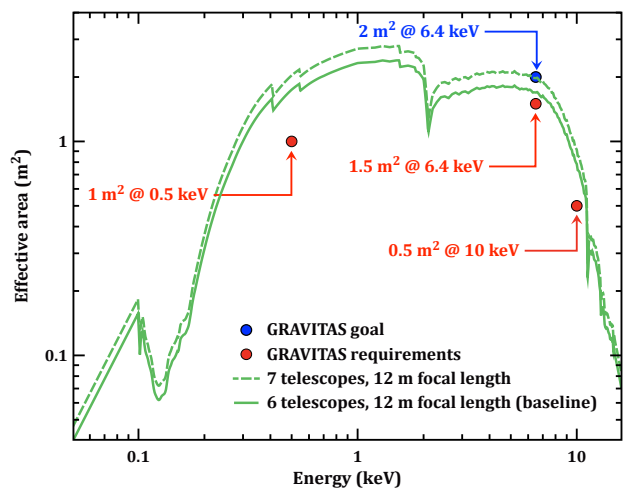

*Figure 3: GRAVITAS effective area curves for the baseline configuration with 6 telescopes and for a configuration with 7 telescopes. The six telescope configuration meets all the requirements, while the 7 telescope option would potentially meet the goal effective area at 6.4 keV.*

GRAVITAS is made of four modules: the Telescope Assembly Module (TAM), the SerVice Module (SVM), the Deployment Module (DM), the Instrument Module (IM) (Figure 4 & Figure 5). The TAM comprises the 6 telescopes and their supporting structure, minimizing losses on the effective collecting areas. The TAM is mounted on top of a support structure under the launcher fairing, and is deployed after launcher separation to obtain the 12 m focal length. A baffle shields the telescopes from the Sun (located at 90+/-20° from the spacecraft main axis). A dedicated deported thermal control unit ensures the thermal regulation of the mirrors, far from the power regulation unit located in the SVM. The mirror temperatures should be kept at 20° ± 1° C. The telescope assembly is covered with multi-layer insulation blankets (MLI). The solar panels are attached to the telescope assembly. During the sun phases, the rear side of the panels supports excessive cooling of this assembly.

The SVM integrates all the satellite functions (propulsion, avionics, communication, and power generation for the SVM, TAM and DM). SVM lateral panels are used as radiators, mutually connected via heat pipes to cool the focal plane of the payload camera heads. The DM consists of the mechanisms, which control the separation of the telescope assembly from the SVM and the IM together; its heritage relies on the IXO studies from both TAS and ASTRIUM. The booms are folded against the SVM in the stowed configuration. The IM comprises the 6 HIFI camera heads operating at the prime focus of the X-ray telescopes, protected from straight light by one meter long baffles, and the associated electronic units. A shield protects the instrument module and SVM panels from the Sun, with the same orientation as the TAM baffle. After deployment of the DM, the IM and the SVM remain attached, greatly simplifying their power and data interfaces.

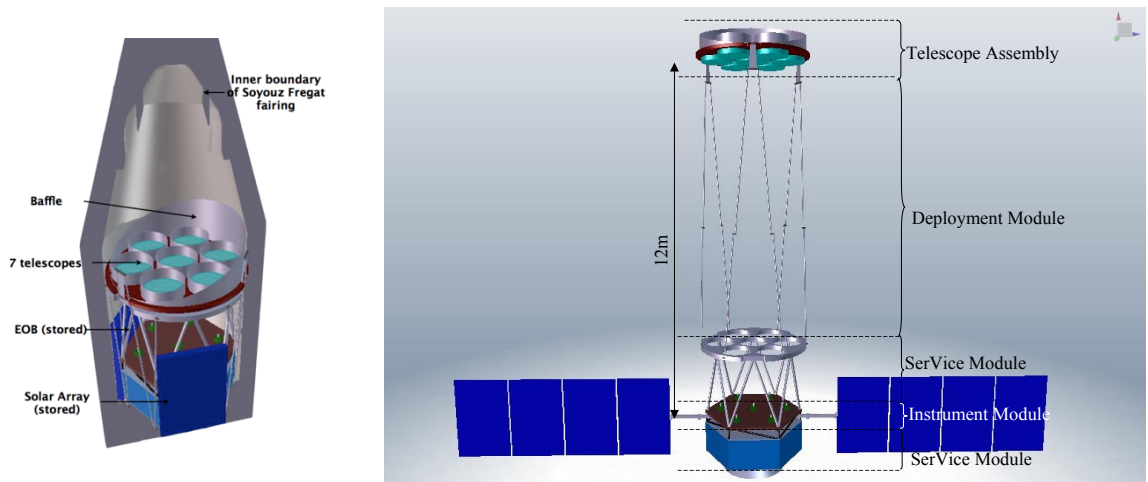

*Figure 4: View of the spacecraft configuration as proposed by the TAS industry study. Left) Spacecraft in stowed position. The seven telescopes of the TAS configuration can be seen. The spacecraft fits within the inner boundary of the Soyuz Fregat fairing. (Right) GRAVITAS in deployed configuration. The extended EOB can be seen. The Service Module (SVM) and the solar arrays (SA) are situated at the instrument module (IM).*

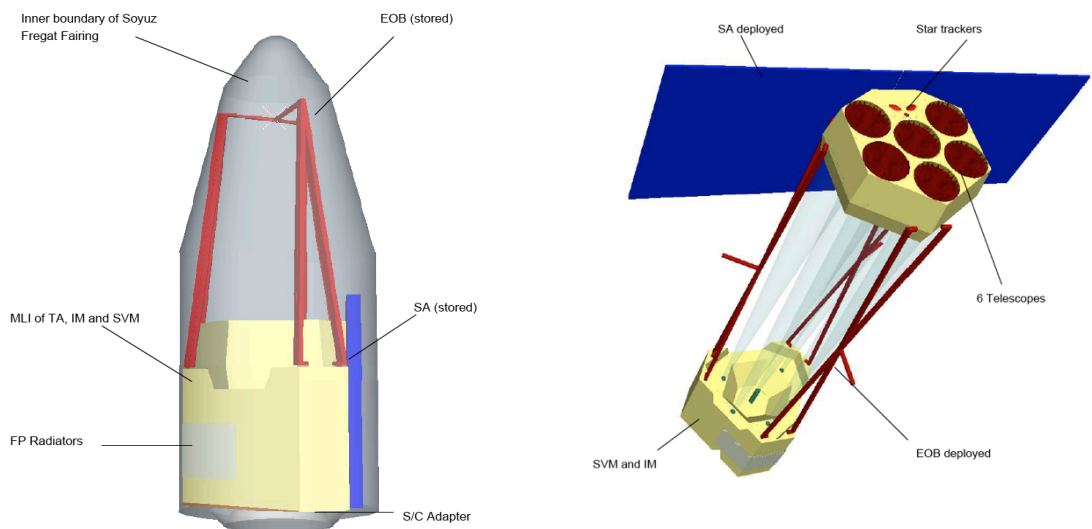

*Figure 5: View of the spacecraft configuration as proposed by ASTRIUM. Left) Spacecraft in stowed position. The spacecraft fits within the inner boundary of the Soyouz Fregat fairing. Right) GRAVITAS in deployed configuration with extended EOB. The ASTRIUM configuration contains six telescopes, leaving more room for metrology at the Telescope Assembly. The SVM is located at the IM, the SA are located at the mirror module and act as a sun-shield.*

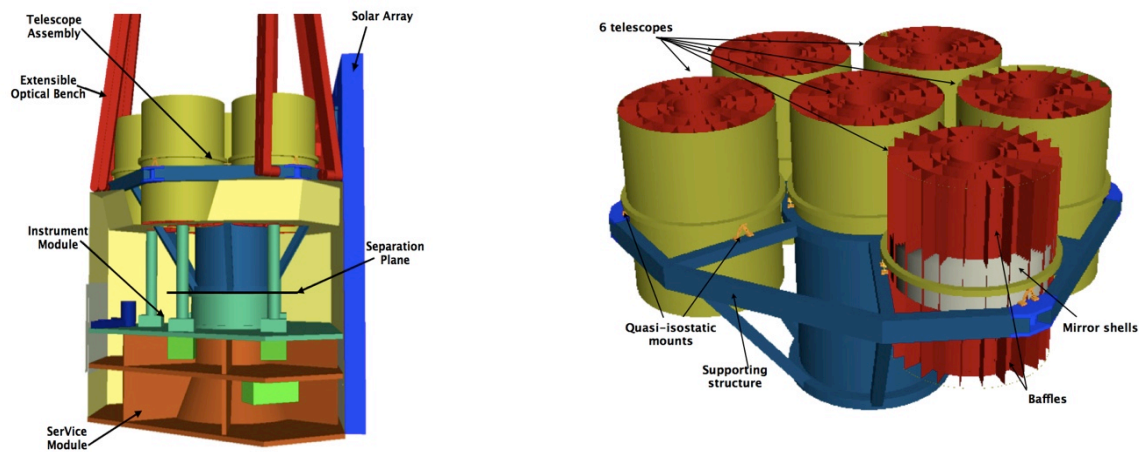

*Figure 6: Left) Internal view of GRAVITAS in launch configuration. Right) Telescope Assembly (TA). 6 telescopes are mounted on a supporting structure, which can distribute the forces during launch onto the central cylinder. The nearest telescope is shown with the outer cylinder removed. The mirror is mounted with 3 quasi-isostatic mounts onto the supporting frame.*

The estimated spacecraft launch mass estimated by ASTRIUM is 4960 kg including approximately 20 % maturity margin, the estimated mass of TAS is 4780 kg.

As can be seen from Figure 6, the optical bench comprises the struts of the extendable optical bench and has no closed shroud, which could prevent interaction with the environment (a deployable shroud was considered to be a high risk item). As consequence the stray-light could enter into the telescope and heat exchange would significantly change. These interactions are reduced by the use of stray-light and thermal baffles on either side of the mirror modules, being about 1 meter long. The impact on the thermal budget of the focal plane is of less importance.

Thermal control of the mirror modules and cameras requires special attention due to the LEO with high thermal load from Earth and albedo and the continuous variation thereof by attitude change. The requirements and solutions are different for the cameras and for the mirror modules. While the thermally passive mirror modules with their large aperture need a stable temperature, the power dissipating cameras need a stable lower temperature, although embedded in a hot thermal environment.

The mirror modules need to be maintained at about 20°C with a tolerance of about 1°C. The mirrors radiate to space with a large aperture and, although the mirror surfaces have a low thermal emissivity, the nesting of the mirrors with intermediate cavities increases the effective emissivity. This is characteristic of Wolter type telescopes and mitigation requires thermal baffles to reduce the field of view to space for thermal radiation and by compensation of the thermal losses by heating of the mirror support structure (spider) since direct heating of the mirrors is not practical. This concept is well proven by XMM-Newton. Detailed sizing of the thermal baffles, heaters and X-ray baffle including the selection of appropriate materials for thermal coupling or de-coupling would be needed, following the XMM-Newton experience.

The telescope structure carrying the mirror modules is designed such as to minimize its thermal losses (by use of MLI) and to distribute the temperature uniformly over the entire structure. Special means are envisaged in order to make use of the heat dissipated by the solar panel in the ASTRIUM approach, e.g. by use of high conductive material and/or heat pipes. Preliminary investigations have shown that this could reduce the heater power requirements from 3 kWatts (kW) to 2 kW.

The 6 detectors need to be cooled to and maintained accurately at a temperature of -50°C with an active heat dissipation of about 24 W in total. This power is transferred to a heat sink, i.e. radiators, using heat pipes. Considering realistic thermal couplings and parasitic heat (eROSITA heritage), the temperature of the radiators must be equal or less than -65 °C. Due to the orbit parameters and attitude needed for the observation, this temperature cannot be reached using a single radiator. Therefore two radiators are accommodated at two opposite sides of the spacecraft, so that at least one of the two radiators has a reasonable view factor to space. Simulations have been performed considering a 600 km orbit. In order to ensure that at least one of the two radiators is at the required temperature and is able to reject the heat, a radiating surface of 3 m² each is needed.

## 2.2  X-ray optics

The GRAVITAS requirements must be met within the mass and geometry constraints imposed by a Soyuz launcher. The driver is the area of 1.5 m² at 6.4 keV which implies to utilise a large fraction of the open diameter of the fairing (~3.8 m) efficiently, packing together a large number of nested grazing incidence mirrors of high reflectivity at 6.4 keV with minimum obstruction from support elements.

The X-ray reflecting coating is iridium. We have considered multilayer coatings but although these can be used to improve the high-energy tail of the response above 10 keV they cannot provide a broad energy band response below 10 keV targeted at the crucial 6.4 keV band. The reflectivity of iridium at 6-7 keV drops rapidly for grazing angles greater than ~0.65 degrees and, for a Wolter I telescope geometry the focal ratio, F/D, is 1/2 tan(4$\alpha$) where $\alpha$ is the grazing angle so we must use a nested Wolter I design with focal ratio F/D>10. Each telescope has an aperture diameter of 1.050 m and a focal length of 12 m so that all the nested shells contribute significantly and efficiently to the area at 6.4 keV.

### 2.2.1  Segmented Slumped Glass Optics

The baseline mirror technology is Segmented Slumped Glass Optics (SSGO), which were originally developed for the High Energy Focusing Telescope (HEFT) balloon payload experiment and are currently being used for the construction of the NuSTAR high energy X-ray telescope mirrors. Their capabilities and development status are very well matched to the GRAVITAS requirements of large collecting area at 6.4 keV and modest angular resolution, ≤1 arc minute. The mirror angular resolution specification for NuSTAR is 45 arc seconds HEW.

The SSGO concept for GRAVITAS is founded on the indirect glass slumping technique pioneered at MPE. Sheets of D 263 T borosilicate glass of thickness ~0.4 mm are formed into a complete parabola and hyperbola Wolter I segments using porous ceramic moulds. Positive moulds are employed so that the reflecting surface is not in contact with the moulds during slumping and the surface roughness of the reflecting surfaces, which is ~0.3 nm rms for the D 263 T glass, is not compromised by the process.

The preferred mould material is a sintered ceramic consisting mainly of aluminum oxide. This has no chemical reaction with the glass so there is no sticking and it is stable during many thermal cycles. It also has a thermal expansion coefficient reasonably well matched to that of the glass.

Because the resulting segments are integrated parabola and hyperbola pairs with an angle 3α (3 times the grazing angle) at the join plane, the moulded sheets have intrinsic stiffness and there is no requirement to subsequently align the surfaces during integration of the segments into a complete nest. The baseline telescope module parameters are:

- Slumped borosilicate glass, d = 0.4 mm, density ρ = 2.51 g/cm², Glass mass per module: 145 kg
- Number of nested Wolter I shells: 242 per telescope module, reflecting coating: iridium
- Outer shell radius $R_{out}$ = 525 mm & inner shell radius $R_{in}$ = 150 mm
- Axial mirror length = 300 mm

If we assume that the required support structure blocks ~10% of the aperture between $R_{in}$ and $R_{out}$ and we suffer scattering and alignment losses of ~4%, then the effective area at 6.4 keV is 1.75 m² for 6 modules which comfortably meets the requirement.

A conceptual design of the mirror modules is presented in Figure 7. The mirror module comprises 242 Wolter I mirrors. This design comprising a total number of 2896 segments is found to be a good compromise between total number of shells and related manufacturing effort, achievable accuracy, deformation under gravity during integration and stress under launch conditions. The segments are integrated directly to a spoke wheel. Alternatively the segments could be integrated in sub-modules to ensure better co-alignment, if needed. Preliminary analyses and tests have shown, that this design could be acceptable.

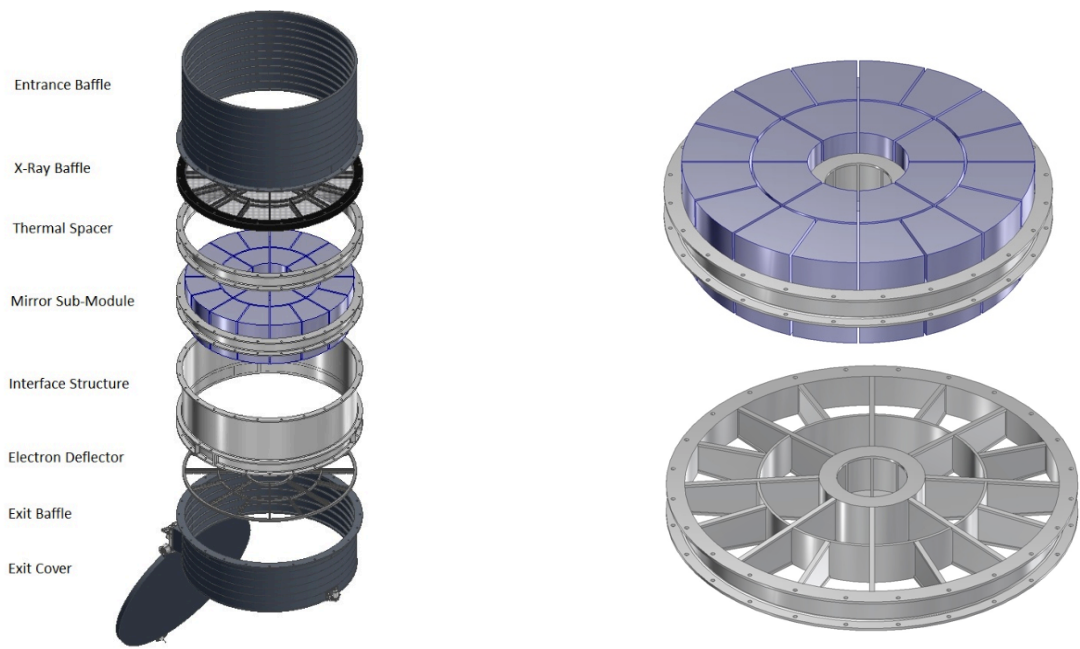

*Figure 7: Left) Exploded view of the complete mirror module, including baffling and interface structure. Top Right) CAD model of glass segments integrated in the mounting structure. Bottom Right) Mirror mounting structure (spider).*

Because the angular resolution requirement for GRAVITAS is 1 arc minute HEW, the demands on the precision, which must be achieved in integrating some 18000 segments into 6 telescope modules is greatly reduced compared with the similar task that was planned for IXO. There are two alternative schemes. The NuSTAR method involves gluing together the shell segments. The alternative approach is to mount each segment using a few points along the axial edges. Because the slumped glass segments incorporate both the parabola and hyperbola surfaces in the MPE indirect slumping technique this point-mounting scheme is considered to be the best approach and has been adopted as the GRAVITAS baseline.

The NuSTAR mirror fabrication was delivering 200 flight segments per week, after the establishment of the manufacturing process, with a total of 8000 modules needed for the mission. For GRAVITAS about 18000 segments would be required leading to a total fabrication time of approximately two years after the end of the development phase.

### 2.2.2 Silicon Pore Optics

The Silicon Pore Optics (SPO) under development for IXO by ESA and Cosine Research in the Netherlands provide a markedly different way of constructing large area and high angular resolution X-ray telescope optics. The driving requirement for this new type of optics was to minimize mass and volume with respect to the large aperture needed for IXO. SPO are produced using commercially available 12 inch Silicon wafers with surface roughness and flatness commensurate with the requirements for the X-ray reflecting surfaces. The wafers are diced, ribbed, wedged and welded together in to stacks to form a matrix of rectangular pores. These stacks are very stiff. The sidewalls of the pores provide internal structure to support the reflecting surfaces and hold them in the required figure for the Wolter I geometry. Two stacks are integrated together to form a High Precision Optic (HPO) module, which acts as an ´X-ray lens´. Each HPO module provides focusing to arc second precision (IXO requirement) and a collecting area of a few cm². Several thousand such HPO modules must be integrated into a large aperture (for IXO sub-divided into several petals) to achieve a large collecting area.

GRAVITAS can benefit from this development, providing mass saving compared to the glass segments. A first estimation from Cosine suggests a total mass of 113 kg for the HPO modules while the proposed baseline for the glass solution amounts to 154 kg, both numbers including 10% margin. Further mass saving is expected to result from lower sensitivity of the SPO to thermal gradients and correspondingly reduced baffling and heating requirements. For GRAVITAS the radius of curvature required at the inner edge of the HPO module is 150 mm while for IXO this inner radius is 250 mm. It is expected that such HPO plates of sufficient accuracy can be produced with the radius of curvature required for the inner regions of the GRAVITAS apertures, but this has not been demonstrated within the IXO SPO development program. On the other hand the angular resolution requirement for GRAVITAS, HEW < 1 arc minute, is greatly relaxed compared with the IXO goal specification, 5 arc seconds HEW. Assuming that the HPO packing that can be achieved is the same as that used for the current IXO SPO design and including 4% losses from scattering and misalignments, then 6 modules give a total of ~1.5 m² at 6.4 keV, just within requirement. It is likely that a more compact packing scheme could be adopted for the HPOs in which case the effective area that could be achieved would be substantially increased.

## 2.3  Focal plane assembly

The GRAVITAS focal plane consists of six identical HIFIs. The DEPFET, a combined detector-amplifier structure, is the central element of HIFI. Every pixel consists of a MOSFET, integrated onto a fully depleted silicon bulk. With an additional implantation, a potential minimum for electrons, the internal gate, is generated below the transistor channel. Electrons generated by the absorption of radiation are collected in the internal gate, where their presence influences the conductivity of the MOSFET channel; sensing the MOSFET's current is therefore a measure of the quantity of collected charge.

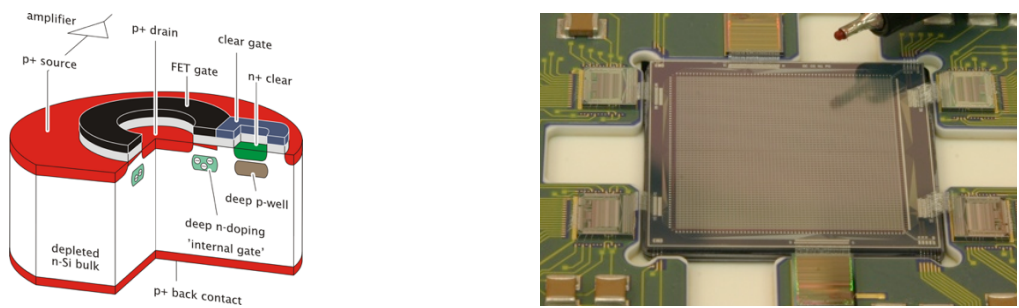

*Figure 8 Left) Cutaway display and schematic of a DEPFET pixel. The pixel can be represented as a combination of a dual-gate p-channel MOSFET and the parasitic n-channel ClearFET structure. (Right) 64 × 64 pixel, 300 µm × 300 µm size (total size: 19.2 x 19.2 mm$^2$) Macropixel detector of the MIXS Lab-Model.*

To achieve a pixel size of 380 µm, HIFI utilizes Macropixels, a combination of the DEPFET (Figure 8) with the Silicon Drift Detector (SDD) concept. A SDD combines a small read-out node with a large sensitive volume without the usually associated large capacitance. By changing the shape and number of drift rings, the size and shape of the SDD can be adjusted freely with only very little effect in the spectroscopic resolution. In a Macropixel, the drift rings of a SDD are combined with a DEPFET as amplifier element, resulting in a detector with adjustable pixel size, excellent energy resolution, low power consumption, and very fast and flexible read-out.

Each HIFI APS consists of a 64 × 64 pixel matrix, integrated monolithically onto a 450 µm thick Si bulk. The pixel size is smaller than the optics resolution element, even the most ambitious goal of 25″ PSF results in an oversampling of 3.8. The field of view achieved by the 2.4 × 2.4 cm² APS size is approximately 7′ for a 12 m focal length.

The camera head contains the base-plate for FPA mounting and cold-finger interface. In the baseline design, both the ceramic board carrying the front-end electronics, as well as the APS itself are kept at detector temperature (-50°C) (see Figure 9). The thermal load can be significantly reduced in a design option in which the APS is thermally de-coupled from the rest of the FPA, and the front-end electronics is kept at S/C ambient temperature. This design would be based on the MIXS thermal concept, but comes at the expense of a more complex FPA. Surrounding the FPA is a graded shield, that is extended towards the instrument baffle. The camera head also contains a filter-wheel with four positions (open, closed, filter and calibrate), which is kept in "filter" position during most observations to mitigate contamination (see Figure 9).

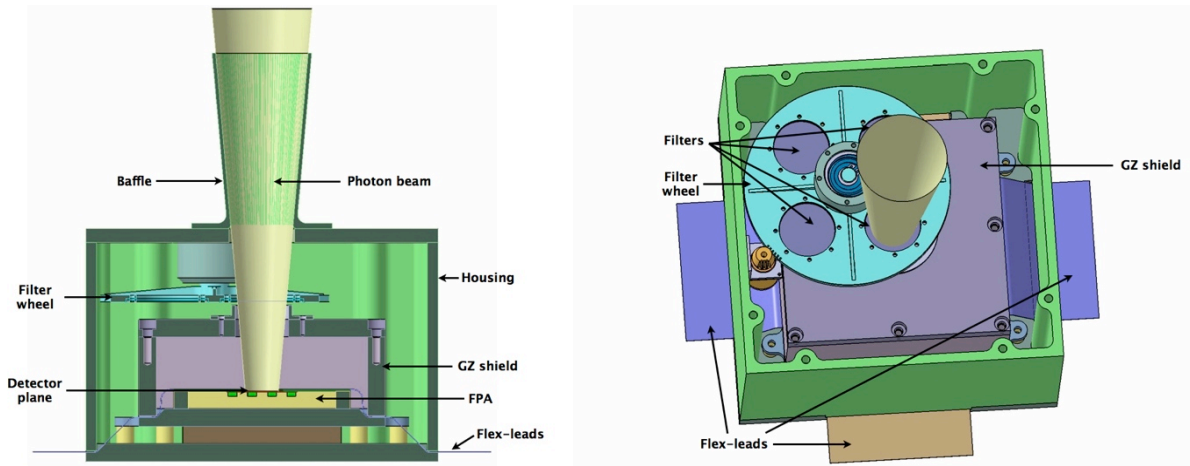

*Figure 9: Left) Cut through a CAD model of the HIFI camera head (baseline thermal concept shown here). The filter wheel, GZ-shield and FPA can be seen. (Right) View into the open HIFI camera head. Flex-leads carrying the analogue signals are indicated in dark blue and thermal I/F in brown.*

The HIFI design meets all GRAVITAS performance requirements. The energy resolution would remain better than 150 eV at 6.4 keV at the end of mission lifetime, by cooling the detector head to -50C (the energy resolution is 125 eV at the beginning of life). The low energy response is optimized by the use of a thin 70 nm Al optical light blocking filter, deposited on the Si bulk. In full-frame mode, HIFI has a frame-time of 64 µs, which can be reduced to 16 µs in a non-destructive 16-line high window mode (two lines per hemisphere are read out simultaneously). The high read-out speed of HIFI allows observation of very bright sources with little pileup and high throughput. End-to-end Monte-Carlo simulations assuming a goal PSF of 25″ have shown that a source of a 150000 cps (or 1 Crab) intensity can be observed with less than 2% pile-up in window mode (with rejection of multiple events, i.e. events depositing signal charge in more than one pixel, so called single events, see Figure 10), and can achieve 92% throughput.

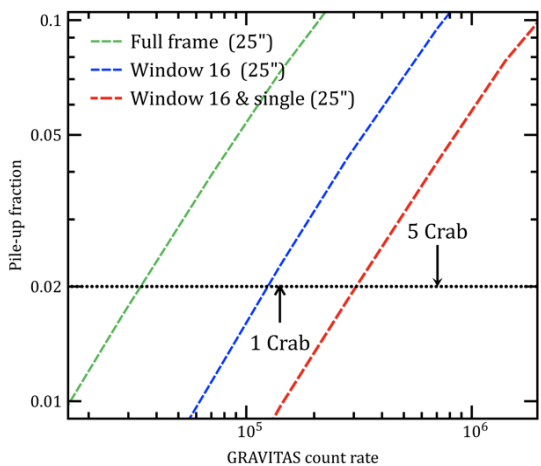

*Figure 10: Pile-up performance of the HIFI instrument. The figure shows the GRAVITAS pile-up as a function of count rate for the different readout modes: the full frame mode (64 µs) and the windowing mode (16 µs). The windowing mode with single events only discards signals, which are shared among two or more pixels.*

Each camera-head assembly is of 20 × 20 × 33.5 cm$^3$ size, and two system electronics box of 25 × 20 × 30 cm$^3$ size. Overall mass of HIFI is 149.5 kg (w/o margin), overall power is 253.7 W (incl. DC-DC loss, w/o margin). The power dissipated at detector temperature (-50°C) is 24.5 W in the baseline configuration, but can be reduced to 3.7 W at the expense of FPA complexity.

The raw internal data rate produced by the HIFIs is 6 × 96 MByte/s. This data rate is efficiently reduced by utilizing the inherently parallel nature of the data stream to a science data stream. In full event mode, the science data rate ranges from 1.6 MBit/s for faint sources to 6 Mbit/s for bright sources in the 100000 cps regime. For maximum count-rate capability, non-essential data are discarded from the data-stream, resulting in a maximal data-rate of 23 Mbit/s for extremely bright (1Mcps regime) X-ray flares.

HIFI offers several operation modes designed for different observational scenarios, all of which can be operated in full-frame or in window mode. In Engineering mode, full raw frame data are downlinked for engineering and creation of calibration data. For sources below about 1-2 $10^5$ counts/s, full event data can be downloaded with lossless compression. For brighter sources, information non-essential to the science objectives can be discarded (e.g. position information, and ADC resolution surpassing the intrinsic energy resolution).

## 3 Mission profile

The scientific requirements for GRAVITAS can be met by a variety of potential orbits. Given the GRAVITAS mass estimate, viable launch options in the context of an M-class mission are restricted to direct injection into a low earth orbit which can be achieved within the capabilities of a Soyuz Fregat launcher (launch capability of ~5.5 tons assuming a launch from Kourou). The baseline orbit has an altitude of 550-600 kilometers and a period of 95-96 minutes with ~0° inclination. This provides stable and low background conditions, avoiding almost entirely the South Atlantic Anomaly (SAA). Similarly the total ionizing dose is small, of the order of $10^5$ 10 MeV equivalent protons after 5 years of operations. This orbit choice also maximizes contacts for suitably located ground stations.

The proposed nominal GRAVITAS scientific operation phase is 3 years, with an extended operational lifetime of 5 years. Normal GRAVITAS science observations are pointed observations of selected targets with observation durations ranging from ~5 ksec to more than one day. The requirement to maximize the area of the sky accessible (and the duration of the visibility period) is met by a design that allows the pointing position to lie within 90°±20° degrees with respect to the solar vector. This provides an instantaneous accessible area of 34% of the celestial sphere and minimum visibility periods (for a specific target) of at least 2.7 months. For the nominal baseline orbit, the observing efficiency of GRAVITAS for the accessible part of the sky ranges from 40% to 80% (depending on target elevation from the orbital plane), taking into account an Earth limb avoidance angle of 25° required to avoid excessive optical stray light expected with the current baffling approach. During science operations, orbit determination is conducted with the required accuracy to meet the observatory timing accuracy requirement. The on-ground a posteriori reconstructed absolute timing accuracy of GRAVITAS is required to be 10 μsec. This is readily available through, e.g., a GNSS receiver.

For downlink, the typical GRAVITAS data volume for one day's observations is 100-300 Gbit and up to 1500 Gbit for special observations of bright targets (anticipated to comprise less than 5% of the observing program). X-band communications can provide 300Mbit/s. A nominal 1-3 telemetry communication periods per day would provide the required downlink capacity for typical observations, assuming the contact time is in the range 6-8 minutes per contact. Additional contact periods may be required to support the data volumes from special observations of bright targets. Alternatively the downlink of the data from such observations could be carried out over a more extended time period. S-band communication is sufficient to support spacecraft command uploads at a frequency of 1-2 times per day. Due to its low earth orbit and due to the inertial attitude of GRAVITAS, the required X-band downlink requires a 2-axis steerable X-band antenna. For the baseline orbit proposed, communications with the spacecraft require a ground station close to the equator. Kourou and the non-ESA station Malindi are optimally placed, assuming the required facilities would be available in the operational time-frame of GRAVITAS.

GRAVITAS has no unusual ground system requirements for its operations. These can be met with a standard ESA ground segment approach for commanding and control of the spacecraft and instruments, monitoring of spacecraft and instrument safety and the downlink of housekeeping and scientific data. The GRAVITAS Mission Operations Centre could be located at ESOC and the Science Operations Centre at ESAC. The GRAVITAS science operations could be organized on relatively conventional lines, similar to those adopted for other ESA missions and adopting some specific features from the approach taken for XMM-Newton and Integral. Mission and science operations could be carried out by ESA through a Mission Operations Centre (MOC) and Science Operations Centre (SOC), together with the Instrument and Science Data Team (ISDT) which is provided by external institutions, funded by national agencies.

## 4   Conclusions

Black Holes and Neutron Stars generate amongst the most extreme physical environments in the Universe. These exotic objects are the locations of the strongest gravitational fields, and the densest matter concentrations, providing unique laboratories to study the limits of General Relativity and fundamental particle physics theories. To determine the behaviour of matter under these extreme conditions, we proposed GRAVITAS (General Relativistic Astrophysics Via Timing and Spectroscopy). GRAVITAS is an X-ray observatory of unprecedented effective area, with a throughput at 6.4 keV around 20 times that of the XMM-Newton EPIC-pn camera. By observing spectral and timing signatures from Active Galactic Nuclei (AGNs) and binary systems on timescales as short as the light crossing time at the event horizon, GRAVITAS can effectively map the strong gravity environment of black holes and neutron stars on the nano-arcsecond scale. The enormous advance in spectral and timing capability compared to current observatories enables a vast array of additional scientific investigations, spanning the entire range of contemporary astrophysics from stars to cosmology.

The model payload consists of six identical telescopes feeding six identical high frame-rate silicon imagers based on DEPFET Active Pixel Sensor technology. The baseline telescope design uses slumped glass coated with iridium, which provides extremely high effective area performance per unit mass. Sub-arcmin angular resolution enables spectroscopy of faint AGNs and galaxy clusters to moderate redshifts. This should be straightforward to achieve and has already been demonstrated by NuSTAR. The required focal length of 12m is achieved by an extendable optical bench. The detectors are state-of-the-art, but none the less have high technology readiness level. They combine near Fano-limited energy resolution (~125 eV FWHM at 6 keV), needed to resolve key atomic features in the X-ray spectra, with the capability to cope with extremely high count rates generated in compact binary systems.

Potential spacecraft configurations have been the subject of two industrial studies. Both show that the mission fits within the available resource envelope in terms of mass, power. GRAVITAS could be launched via direct injection into a 600km equatorial orbit, giving low background. The nominal lifetime is 3 years, with possible extension to 5. Ground segment and operations are standard and require no special constraints.

With GRAVITAS, we proposed a mission with focused scientific goals addressing key questions about the nature of black holes and neutron stars, among the most exotic objects in the universe, with implications for basic physical theories. This is a subject no only of wide interest within the astrophysics community, but one also able to capture the imagination of the public, who hold a deep and enduring fascination with these extreme phenomena.

# 5 Acknowledgements

It is a pleasure to thank the following people for their contribution to the proposal: Monique Arnaud, **Xavier Barcons**[1], **Werner Becker**, Giovanni Bignami, Mitch Begelman, Sudip Bhattacharyya, **Hans Böhringer**, Slavko Bogdanov, **Thomas Boller**, **Martin Boutelier**, Niel Brandt, Heinrich Bräuninger, **Marcella Brusa**, Herman Brunner, **Ed Cackett**, **Massimo Cappi**, Francisco Carrera, Pablo Cassatella, Sandip Chakrabarti, Sylvain Chaty, **Andrea Comastri**, Thierry Courvoisier, Mauro Dadina, Anne Decourchelle, Melania Del Santo, Konrad Dennerl, Maria Diaz Trigo, Chris Done, Tadayasu Dotani, Jeremy Drake, **Josef Eder**, Sean Farrell, **Alexis Finoguenov**, George Fraser, **Michael Freyberg**, **Peter Friedrich**, **Luigi Gallo**, Ian George, Marat Gilfanov, Margherita Giustini, Olivier Godet, Ersin Gogus, Nicolas Grosso, **Frank Haberl**, **Fiona Harrison**, Coel Hellier, Vladimir Karas, Wlodek Kluzniak, Stefanie Komassa, **Peter Lechner**, Roy Lemmon, Dangbo Liu, Dipankar Maitra, Julien Malzac, Hironori Matsumoto, Ian McHardy, Mariano Mendez, **Andrea Merloni**, **Cole Miller**, **Jon Miller**, **Richard Mushotzky**, Andreas Mueller, Jean-François Olive, Manfred Pakull, Stéphane Paltani, Iossif Papadakis, Joachim Paul, Pierre-Olivier Petrucci, Wolfgang Pietsch, Etienne Pointecouteau, Gabriele Ponti, Delphine Porquet, **Juri Poutanen**, Katja Pottschmidt, **Dimitrios Psaltis**, Arne Rau, Alak Ray, Martin Rees, **Chris Reynolds**, **Ruben Reis**, Thomas Reiprich, Guido Risaliti, Bob Rutledge, Andrea Santangelo, **Christian Schmid**, Jurgen Schmitt, Nikolai Shaposhnikov, , Roberto Soria**, Alexander Stefanescu**, Gordon Stewart, Raschid Sunyaev, Michel Tagger, **Tadayuki Takahashi**, **Francesco Tombesi**, Christof Tenzer, Hiroshi Tsunemi, **Phil Uttley**, Simon Vaughan, Dominic Walton, Bob Warwick, Natalie Webb, Klaus Werner, Pete Wheatley, **Joern Wilms**, **Anita Winter**, Wenfei Yu, Ye-Fei Yuan, Silvia Zane, Andrzej Zdziarski, **Abdu Zoghbi.**

We are grateful to Astrium and Thalès Alénia Space for their engineering support along the preparation of the GRAVITAS proposal.

---

[1]Main contributors to the proposal are indicated in bold face.